\documentclass[aps,prl,twocolumn,groupedaddress,latterpaper,preprintnumbers,amsmath,amssymb]{revtex4}
\usepackage{graphicx}
\usepackage{dcolumn}

\begin{document}

\title{Effects of localization and amplification on distribution of intensity transmitted through random media}
\author{Alexey Yamilov and  Hui Cao}
\affiliation{Department of Physics and Astronomy\\ Northwestern University, Evanston, IL 60208}
\email{a-yamilov@northwestern.edu}

\date{\today}

\begin{abstract}
We numerically study the statistical distribution of intensity transmitted through quasi-one dimensional random media by varying the dimensionless conductance $g$ and the amount of absorption or gain. Markedly non-Rayleigh distribution is found to be well fitted by the analytical formula of Nieuwenhuizen {\it et al}, Phys. Rev. Lett. {\bf 74}, 2674 (1995) with a single parameter $g^\prime$. We show that in the passive random system $g^\prime$ is uniquely related to $g$, while in  amplifying/absorbing random media $g^\prime$ also depends on gain/absorption coefficient.

\end{abstract}


\maketitle

Light transport in a quasi-one dimensional system is described by three directly measurable quantities: $T_{ab}$ -- transmission coefficient from an incoming channel $a$ to an outgoing channel $b$; $T_a=\sum\limits_b{T_{ab}}$ -- total transmission coefficient from channel $a$ to all outgoing channels; $\tilde{g}=\sum\limits_{a,b}{T_{ab}}$ -- transmittance\cite{altshuler,feng_and_vanrossum}. A wave transmitted through a random medium is a coherent sum of a large number of contributions due to different propagation paths. In the diffusive regime, $\langle \tilde{g}\rangle \equiv g\gg 1$, the contributions $T_{ab}$ are largely uncorrelated. Gaussian distribution of the field components leads to renowned Rayleigh distribution $P(s_{ab})=\exp{(-s_{ab})}$ of the normalized intensity\cite{outside_intensity}, $s_{ab}=T_{ab}/\langle T_{ab}\rangle$, with a simple relation between the moments of the distribution: $\langle s_{ab}^n\rangle=n!\langle s_{ab}\rangle^n$. Such factorization cannot be exact\cite{sherb_kaveh}, since it implies the complete absence of nonlocal correlations\cite{nonlocal}. Kogan {\it et al} \cite{kogan_first} related the distribution of the normalized total transmission $s_a=T_{a}/\langle T_{a}\rangle$ to that of $s_{ab}$:
\begin{equation}
P(s_{ab})=\int\limits^{\infty}_0 \frac{ds_{a}}{s_{a}} P(s_{a}) \exp{\left[-\frac{s_{ab}}{s_{a}}\right]} .
\label{pofi}
\end{equation}
The Gaussian distribution of $P(s_{a})$ for $g\gg1$ leads to the deviation from Rayleigh distribution at large intensity $s_{ab}\gg g$. The deviation has the same origin as the universal conductance fluctuations and nonlocal intensity correlations in transport\cite{altshuler,feng_and_vanrossum}. The distribution of the intensity was obtained in a closed form in Refs. \onlinecite{van_rossum,kogan_kaveh}
\begin{equation}
P(s_{a})=\int\limits_{-i \infty}^{i \infty}\frac{dx}{2\pi i}
\exp{\left[xs_{a}-\Phi_0(g^{\prime},x)\right]},
\label{poft}
\end{equation}
where $\Phi_0(g^{\prime},x)=g^{\prime}\ln^2\left( \sqrt{1+x/g^{\prime}}+\sqrt{x/g^{\prime}}\right)$. Eq. (\ref{poft}) was derived under the assumption of $g\gg 1$, thus $g^\prime\equiv g$. The expressions for $P(s_{a})$ and $P(s_{ab})$ have been verified in experiments \cite{marin_total_transm,marin_intensity_distribution,azi_nature,azi_in_sebbah}. Unexpectedly, the experiments demonstrated that Eq. (\ref{poft}) worked well even for moderate values of $g\sim 10$, and in the presence of significant absorption. Moreover, based on the statistics of the transmitted intensity, the localization criterion, 
\begin{equation}
g^\prime\equiv 2/3{\rm var} (s_{a})\equiv 4/3[{\rm var} (s_{ab})-1]
\label{gprime}
\end{equation}
equals to unity, was surmised\cite{azi_nature}. Such definition of $g^\prime$ can be used, irrespective if the Eq.(\ref{poft}) holds.  However, if Eq.(\ref{poft}) is applicable, $g^\prime$ obtained from  Eq.(\ref{gprime}) should match the one obtained from the fit of  the entire distribution of $s_{ab}$ with Eqs.(\ref{pofi},\ref{poft}) \cite{kogan_kaveh}. One question to be addressed in this paper is whether the Eqs.(1,2) still hold in the regime of incipient photon localization $g \sim 1$, and if so, how the intensity distribution fitting parameter $g^\prime$ is related to the system properties.

We will also investigate the effects of optical amplification and absorption on the transmitted intensity distribution. The effect of the amplification on purely 1D transport was studied by a number of groups \cite{1d_distributions}. In a quasi-1D system, the probability of reflectance by a random amplifying medium was also obtained by Beenakker {\it et al}\cite{beenakker}. 
Zyuzin showed that the fluctuation of $T_a$ grows faster than its average value in a amplifying random medium near the lasing threshold\cite{zyuzinPRE}. However, it is not clear whether the statistical distribution of transmission coefficient of an active system differs qualitatively from that of a passive system. Brouwer demonstrated that Eq.(\ref{poft}) is inapplicable in the limit of strong absorption \cite{brouwer,pnini_in sebbah}, whereas we are interested in the weakly absorbing systems with $\sqrt{D\tau }> L$, where $D$ is the diffusion coefficient, and $\tau$ is the absorption time. 

In this paper, we numerically calculate the statistical distribution of transmitted intensity $P(s_{ab})$ in a quasi-1D system. In particular, we check whether $P(s_{ab})$ still satisfies Eqs.(1,2). In real experiment or numerical study, the local intensity distribution $P(s_{ab})$ is obtained by collecting the data of transmitted intensity from many random configurations. Among them, there exist rare configurations that could lase in the presence of gain. Light intensity would diverge if gain saturation is neglected.  In the diffusive regime $g \gg 1$, this problem is limited only to the immediate vicinity of the diffusive lasing threshold\cite{zyuzinPRE}. For the systems we considered, $g < 10$. Strong fluctuation of lasing threshold results in a non-negligible percentage of realizations whose lasing threshold is met even at moderate gain (i.e., not very close to the diffusive lasing threshold). Although gain saturation could prevent the divergence of laser intensity, the actual value of the saturated intensity depends on the properties of the gain material. In order to eliminate any material-dependent effect on $P(s_{ab})$, we disregard the contributions of the lasing configurations to the intensity distribution\cite{our_active_work}. Our numerical studies show that the resultant $P(s_{ab})$ can be well described by Eqs.(1,2) with a reduced $g^\prime$. 

\begin{figure}
\centerline{\rotatebox{-90}{\scalebox{0.3}{\includegraphics{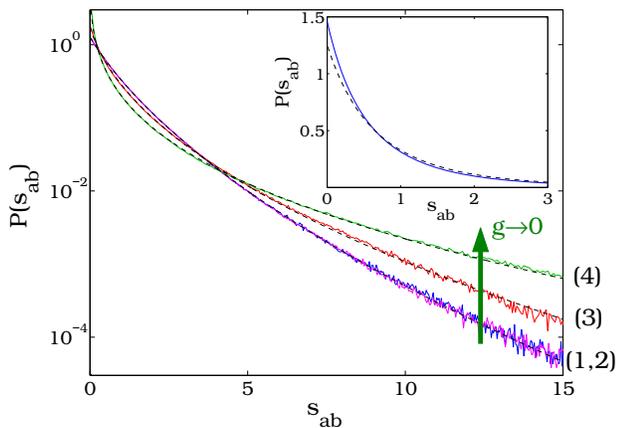}}}}
\caption{\label{p_of_i_passive} Numerically calculated $P(s_{ab})$ for four samples. Dashed lines represent the fit with Eqs.(\ref{pofi},\ref{poft}). The inset shows $P(s_{ab})$ calculated from Eqs.(\ref{pofi},\ref{poft}) with the kernel $\Phi_0(g^{\prime},x)$ (solid line), or $\Phi_1(g^{\prime},x)$ (dashed line). $g^\prime=3$.}
\end{figure}

We numerically obtain $P(s_{ab})$ as the histogram of the intensity transmitted through a quasi-1D system -- a 2D metallic waveguide filled with circular dielectric scatterers. We employ the method developed earlier in Ref. \onlinecite{our_passive_work} for passive and Ref. \onlinecite{our_active_work} for amplifying and absorbing systems. In quasi-1D geometry the transition from diffusion $g\gg1$ to localization  $g\lesssim1$ can be achieved even in the  weak scattering regime by increasing the system length $L$  beyond the localization length $\xi$. First, we study the intensity distribution in passive systems as $g$ approaches $1$.  The random system is characterized by the Thouless number $\delta=\delta \nu/\Delta\nu$. The average mode linewidth $\delta \nu$ is obtained from the width of $|C_E(\Delta \nu)|^2$ divided by a numerical factor $1.46$ ($C_E(\Delta \nu)$ is the spectral field correlation function). The average mode spacing is $\Delta \nu=c/\left(\pi L^\prime N n_{eff}^2\right)$, where $L^\prime=L+2z_b$ is the effective length of the random media, $z_b$ accounts for the boundary effects. $N$ is the number of waveguide modes, $c$ is the speed of light in vacuum, and $n_{eff}$ is the effective refractive index. All these parameters can be determined independently\cite{our_passive_work,our_active_work}. In passive diffusive system $\delta$ coincides with the dimensionless conductance $g$. We consider four samples labeled from 1 to 4. The lengths of the first two samples are related as $L_2=2L_1$, and scatterer density is chosen in such a way that the increase of $L$ would be offset by the change of the transport parameters, mostly by the transport mean free path $l$, to give the identical value of  $\delta=4.4$. The samples $\#3$ and $\#4$ had the same scatterer densities as in $\#1$, but $L$ is increased yielding $\delta=2.2$ and $\delta=1.13$ respectively.
Fig. \ref{p_of_i_passive} shows that samples $\#1$ and $\#2$ are fitted by Eqs.(\ref{pofi},\ref{poft}) with the same $g^\prime=2.9$. This means that the distribution of local intensity is determined by single parameter $g^\prime$, which depends only on $\delta$ (or $g$, see below). The samples $\#3$ and $\#4$ are fitted well with $g^\prime =1.25$ and $g^\prime =0.4$. Such agreement confirms the applicability of Eqs.(\ref{pofi},\ref{poft}) down to $g^\prime =0.4$, even smaller than that reported in microwave experiments of Refs. \onlinecite{marin_intensity_distribution,azi_nature,azi_in_sebbah}. In what follows we give an argument that may shed some light on this agreement.

\begin{figure}
\centerline{\rotatebox{0}{\scalebox{0.35}{\includegraphics{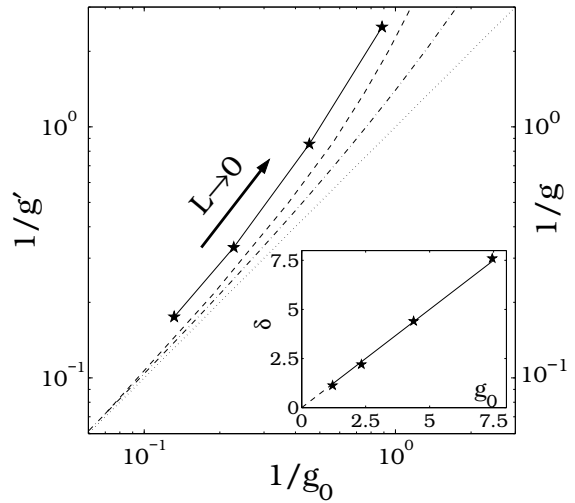}}}}
\caption{\label{gprime_vs_g_passive} Symbols depict $1/g^\prime$ versus $1/g_o$ for samples $\#1,3-5$. In the zeroth order of $1/g_o$, $g^\prime(g_o)=g_o$ (dotted line). The dash-dotted line is $g^\prime(g_o)$ with the first-order correction. Dashed line plots the exact solution $g(g_o)$ from Ref. \onlinecite{mirlin_g}. The inset shows the linear dependence of $\delta$ on $g_o$. The solid line has a slope of one. }
\end{figure}

In Ref. \onlinecite{van_langen} van Langen {\it et al}  found that the proper perturbation series should be constructed for $\Phi(g^{\prime},x)$, the kernel in Eq.(\ref{poft}). The first order (in $1/g$) correction to $\Phi_0(g^{\prime},x)$ of Refs. \onlinecite{van_rossum,kogan_kaveh} was calculated\cite{van_langen}. In the inset of Fig. \ref{p_of_i_passive} we plotted $P(s_{ab})$ calculated within both approximations. We see that Eq. (\ref{poft}) with $\Phi_0$ starts to fail for $s_{ab}< 1$ region, while the asymptotic behavior for $s_{ab}\gtrsim 1$ is preserved. In the transition from diffusion to localization threshold ($g\sim1$), $P(s_{ab}\gtrsim 1)$ should be well described by  Eqs.(\ref{pofi},\ref{poft}) with decreasing $g^\prime$ -- quantitative change. But for $g\ll 1$, $P(s_{ab})$ changes to the lognormal distribution -- qualitative change, not captured by $\Phi_0$ approximation. The numerically obtained intensity distribution, indeed, shows some deviation from Eqs.(\ref{pofi},\ref{poft}) at small values of $s_{ab}$. The deviation, however, is noticeable only when $P(s_{ab})$ is plotted on the linear scale. On the logarithmic scale, the distribution has a pronounced tail at large $s_{ab}$, where the correct asymptotic prevails. 

Next, we would like to analyze the value of $g^\prime$. The following four quantities should coincide in a diffusive system: $g^\prime$, $g$, $\delta$, and $g_0=(\pi/2) n_{eff}Nl/L^\prime$. Fig. \ref{gprime_vs_g_passive} shows the relations of these four quantities in the regime of incipient localization. The numerical data included sample $\#5$ of $\delta=7.6$. The inset of Fig. \ref{gprime_vs_g_passive} shows that for the studied systems $\delta\simeq g_o$, and thus our data for $\delta(g_o)$ should lie along the dotted line of slope $1$ in the main plot. The rigorous solution $g(g_o)$ obtained by Mirlin in Ref. \onlinecite{mirlin_g} is shown in Fig. \ref{gprime_vs_g_passive} with the dashed line. Our data for  $1/g^\prime$ lies above the dotted and dashed lines, revealing the tendency of $g^\prime$ to decrease below $g_o$, $\delta$, and $g$ in the regime $g_o\sim 1$. The first order (in $1/g_o$) correction to $g^\prime (g_o)\simeq g_o$ can be obtained from the variance of the intensity distribution, ${\rm var}{\left(s_{ab}\right)} \simeq 1+4/3g_o+8/15g_o^2$ upto the second order of $1/g_o$\cite{feng_and_vanrossum}. The approximation of $g^\prime(g_o)$ obtained by substitution of the above expression into Eq.(\ref{gprime}) is plotted in Fig. \ref{gprime_vs_g_passive} with the dash-dotted line. The fact that the data points of $1/g^\prime$ is above the dash-dotted line indicates the higher-order terms are not negligible and their contributions lead to a further reduction of $g^\prime$. 

\begin{figure}
\centerline{\rotatebox{-90}{\scalebox{0.3}{\includegraphics{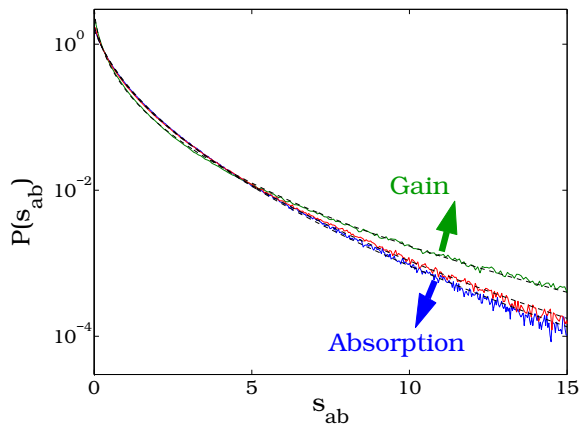}}}}
\caption{\label{p_of_i_ga} Numerically calculated $P(s_{ab})$ for sample $\#3$ with and without gain/absorption. The arrows shows the direction of increasing gain/absorption. Dashed lines represent the fit with Eqs.(\ref{pofi},\ref{poft}) with $g^\prime$ equal to $1.5$ (absorption), $1.25$ (passive), and $0.65$ (gain). }
\end{figure}

Next, we study the effect of coherent amplification on intensity distribution. Fig. \ref{p_of_i_ga} compares the local intensity distributions found in sample $\#3$ with and without gain/absorption. The value of the absorption time is the same as that of the amplification time. The latter was equal to $\tau=5 \tau^{cr}$, where $\tau^{cr}$ is the critical amplification time corresponding to the diffusive lasing threshold\cite{letokhov}. Even at such low level of gain, some of the random realizations lased, due to strong fluctuation of lasing threshold. The numerical results presented in Fig. \ref{p_of_i_ga} contain only the contributions from non-lasing realizations. 

First of all, we would like to point out that Eqs.(\ref{pofi},\ref{poft}) give a good fit to the numerically obtained $P(s_{ab})$ with gain/absorption. In Refs. \onlinecite{marin_intensity_distribution,azi_in_sebbah} it was also found that Eqs.(\ref{pofi},\ref{poft}) describe well the intensity distribution even for absorbing systems. Here, we present a systematic study of the effect of absorption on $P(s_{ab})$ and also consider, for the first time, the effect of amplification. Fig. \ref{p_of_i_ga} shows that the presence of gain leads to an increase of $P(s_{ab})$ in the regions $s_{ab}\ll 1$ and $s_{ab}\gg 1$ and, therefore, an enhancement of intensity fluctuations. The effect of absorption is exactly the opposite. From this qualitative analysis and Eq.(\ref{gprime}), we see that amplification reduces the value of $g^\prime$, while absorption increases it. 

\begin{figure}
\centerline{\rotatebox{0}{\scalebox{0.35}{\includegraphics{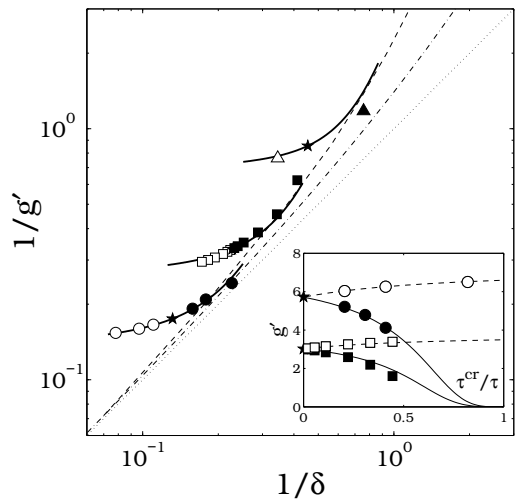}}}}
\caption{\label{gprime_vs_g_ga} The numerical data of $1/g^\prime$ versus $1/\delta$ for samples $\#1$ (squares) , $\#3$ (triangles) , $\#5$ (circles). Solid (open) symbols represent the systems with gain (absorption). The pentagons correspond to the passive system. The dotted, dashed, and dash-dotted lines are the same as in Fig.\ref{gprime_vs_g_passive}. Solid lines represents $g^\prime(\delta)$. The inset shows the explicit dependence of $g^\prime $ on the amplification/absorption time $\tau$ normalized to $\tau_{cr}$. Dashed (solid) line plots Eq.(\ref{as}) with positive (negative) $\tau$.}
\end{figure}

Fig. \ref{gprime_vs_g_ga}  shows the effect of amplification/absorption on  $g^\prime$ for samples $\#1,3,5$. We can still define $\delta$ for amplifying/absorbing systems through the correlation functions\cite{our_active_work}. In contrast to $g_o$, $\delta$ depends on the amount of gain/absorption. The change of $\delta$ is directly related to the change of the correlation linewidth $\delta \nu$, while $\Delta \nu$ remains the same as in passive system. The superlinear increase of $1/g^\prime$ with $1/\delta$ in Fig. \ref{gprime_vs_g_ga} demonstrates that $g^\prime$ decreases significantly faster than $\delta$ with increasing gain. This indicates that intensity fluctuations are more sensitive to amplification than the average mode linewidth $\delta \nu$. The long propagation paths extract more gain, leading to an enhancement of  non-local correlations\cite{our_active_work}. The enhanced correlation results in a further deviation from Rayleigh statistics. The absorption, on the other hand, causes a reduction of fluctuations. The sublinear decrease of $1/g^\prime$ with $1/\delta$ in Fig. \ref{gprime_vs_g_ga} reveals that $g^\prime$  increases slower than $\delta$ with increasing absorption. The dependence of $g^\prime$ on the absorption time $\tau$ can be determined pertubatively from the known result for ${\rm var}(s_{ab})$ in an absorbing system of $g_o\gg 1$:
\begin{equation}
\frac{4}{3 g^{\prime}(g_o,\tau)}\equiv {\rm var}(s_{ab})-1=\frac{4}{3g_o}A_2(\tau)+\frac{8}{15g_o^2}A_3(\tau) \, ,
\label{as}
\end{equation}
where $A_2(\tau)$ and $A_3(\tau)$ are given in Refs. \onlinecite{bing_hu,pnini_in sebbah}. The expressions for $A_i$ were derived for absorbing systems, where $\tau$ is positive. In the negative absorption model, $\tau$ becomes negative for systems with gain. In Ref. \onlinecite{our_active_work} the applicability of the negative-$\tau$ expressions was discussed. Obviously, the contribution of the lasing realizations should be omitted to avoid the divergence of ${\rm var}(s_{ab})$. For the samples $\#1$ and $5$ that have larger $g_o$, we compared our data to Eq.(\ref{as}) in the inset of Fig. \ref{gprime_vs_g_ga}. Also, $\delta (g_o,\tau)$ can be found from the width of $|C_E(\Delta\nu)|^2$\cite{our_active_work}. By eliminating $\tau$ from $g^\prime (g_o,\tau)$ and $\delta (g_o,\tau)$ we obtained $g^\prime(\delta)$, shown as the solid lines in Fig. \ref{gprime_vs_g_ga}. The deviations of our data points from the solid line increase with the decrease of $g_o$, because the contributions of the higher order terms in $1/g_o$ cannot be neglected.

\begin{table}
\caption{\label{tab:table} Comparison of localization threshold criteria in passive system. }
\begin{ruledtabular}
\begin{tabular}{cccc}
$g^\prime$&$g_o$&$\delta$&$g$\\
\hline
1.0 & 2.0 & 2.0 & 1.3\\
0.3 & 1.0 & 1.0 & 0.43\\
0.7 & 1.6 & 1.6 & 1.0\\
\end{tabular}
\end{ruledtabular}
\end{table}
In Ref. \onlinecite{azi_nature} $g^\prime$ was proposed as the localization criterion parameter. Table \ref{tab:table} shows that in passive systems the criteria based on $g^\prime=1$, $g_o=1$, $g=1$, or $\delta=1$ differ by only a numerical factor of order  one. In the presence of gain/absorption the situation is essentially different. By definition, $g_o$ only contains  information of passive system, whereas $g^\prime$ and $\delta$ account for the effect of amplification/absorption. In this case the difference between $g^\prime$, $\delta$, and $g$, is not merely numerical. Amplification/absorption results in a decrease/increase of  $g^\prime$ and $\delta$, but an increase/decrease of $g$. With the increase of absorption, $\delta$ increases without a bound, while $g^\prime(\tau\rightarrow0)\rightarrow 4g^\prime(\tau\rightarrow\infty)/3$ in the limit $g\gg 1$. In sharp contrast, we see that in amplifying systems $g^\prime$ diminishes superlinearly with $\tau^{cr}/\tau$ (inset of Fig. \ref{gprime_vs_g_ga}), while $\delta$ decreases almost linearly with $\tau^{cr}/\tau$ as reported in Ref. \onlinecite{our_active_work}. Our numerical result in the inset of Fig. \ref{gprime_vs_g_ga} also suggests that $g^\prime$ fall below unity prior to $\delta$ in an amplifying system. Therefore,  $g^\prime$ is more sensitive to amplification but less sensitive to absorption than $\delta$. 

In conclusion, we numerically calculated the statistical distribution of transmitted intensity in quasi-1D random medium. In a passive system $P(s_{ab})$ is well described by Eqs.(\ref{pofi},\ref{poft}) down to $g^\prime = 0.4$, far beyond the $g^\prime \gg 1$ regime where it was originally derived. Surprisingly, Eqs. (\ref{pofi},\ref{poft}) also hold for amplifying/absorbing systems with $g^\prime$ different from the value of the passive system. Our data show that $g^\prime$ decreases superlinearly with gain constant, but increases sublinearly with absorption constant. 

The authors acknowledge stimulating discussions with A. Z. Genack and A. Chabanov. This work is supported by the National Science Foundation under Grant No DMR-0093949. HC acknowledges the support from the David and Lucille Packard Foundation.

\end{document}